\documentclass[twocolumn,usenatbib]{mnras}
\usepackage{graphicx}
\usepackage[export]{adjustbox}
\usepackage{   }
\usepackage[colorinlistoftodos]{todonotes}
\usepackage{amsmath}
\usepackage{amssymb}
\usepackage{comment}
\usepackage{multirow}
\usepackage[normalem]{ulem}

\newcommand{\halfskip}{\vskip 0.5\baselineskip}
\newcommand{\be}{\halfskip \begin{equation}}
\newcommand{\ee}{\end{equation} \halfskip \noindent}
\newcommand{\ba}{\halfskip \begin{eqnarray}}
\newcommand{\ea}{\end{eqnarray} \halfskip \noindent}
\newcommand{\auger}{Auger}

\usepackage{subcaption}
\usepackage{setspace}

\begin{document}
\twocolumn

\title{Constraining the Extragalactic Magnetic Field: Auger Data Meet UHECR Propagation Modeling}
\author[A. AL-Zetoun et al.]{
Ala'a AL-Zetoun$^{1,2}$\thanks{E-mail: a.al-zetoun@astro.ru.nl},
Arjen van Vliet$^{1}$\thanks{E-mail: arjen.vliet@ku.ac.ae},
Andrew M. Taylor$^{3}$,
Walter Winter$^{3}$
\\
$^{1}$Department of Physics, Khalifa University of Science and Technology, P.O. Box 127788, Abu Dhabi, United Arab Emirates\\
$^{2}$Department of Astrophysics, IMAPP, Radboud University, P.O. Box 9010, 6500 GL Nijmegen, The Netherlands\\
$^{3}$Deutsches Elektronen-Synchrotron DESY, Platanenallee 6, 15738 Zeuthen, Germany
}

\date{Accepted..., Received...; in original form ...}
\pubyear{2025}

\maketitle
\begin{abstract}

Recent analyses from the Pierre Auger Collaboration suggest correlations between the arrival directions of ultra-high-energy cosmic rays (UHECRs) and catalogs of starburst galaxies (SGBs) and jetted active galactic nuclei (AGNs). We revisit these analyses using the same methodology as \auger , but explicitly incorporating UHECR deflections in turbulent extragalactic magnetic fields (EGMFs). We demonstrate that while for SBGs the same sources as for the generic \auger\ analysis dominate the catalog correlations, jetted AGNs are dominated by Centaurus~A when accounting for source distances and deflections. Using our framework, we derive 90\% confidence level upper limits on the local EGMF strength of 4.4~nG~Mpc$^{1/2}$ for SBGs and 6.7~nG~Mpc$^{1/2}$ for jetted AGNs. Assuming instead that the UHECR deflections predominantly arise from the Galactic magnetic field (GMF), we obtain a GMF upper limit of $1.4 \, \mu$G~kpc$^{1/2}$ for a Galactic halo size of 30~kpc.

\end{abstract}


\section{Introduction}
\label{sec:intro}

Cosmic rays are energetic charged particles  that arrive at Earth with a range of energies spanning many orders of magnitude up to about $10^{20}$~eV. 
Theoretical considerations suggest that ultra-high-energy cosmic rays (UHECRs) are likely produced by extreme energetic events giving rise to high velocity outflows, such as active galactic nuclei (AGNs), gamma-ray bursts (GRBs) or tidal disruption events (TDEs)~\citep{Hillas_1984, Aharonian_2005, RevModPhys.56.255, 2014arXiv1411.0704F}. Sometimes also starbust galaxies (SBGs) are considered \citep[see e.g.][]{2018PhRvD..97f3010A}, although the sub-relativistic velocities of their outflows cast doubt on whether they are able to accelerate particles up to the UHE regime~\citep{2018A&A...616A..57R}.   

The Pierre Auger Observatory (\auger)~\citep{2015172}, in the Southern Hemisphere, and the Telescope Array (TA)~\citep{ABUZAYYAD201287}, in the Northern Hemisphere, are the two largest detectors of UHECRs. \auger\ has recently reported the detection of large-scale anisotropy, namely a dipole in the arrival directions of UHECRs with energies above $8$~EeV with a post-trial significance for a departure from isotropy at the level of $6.8\sigma$~\citep{2017, Aab_2018, PierreAuger:2024fgl}. The fact that this dipole points away from the Galactic center suggests that the majority of UHECR sources at these energies are of extragalactic, rather than Galactic, origin. Furthermore, the latest joint analysis combining data from both \auger\ and TA has reached a post-trial significance for a departure from isotropy at the level of $4.6\sigma$~\citep{TelescopeArray:2025yvu}.

At even higher energies, $\gtrsim 40$~EeV, \auger\ has observed a significant excess of events on smaller angular scales, which appears to correlate with local extragalactic structures. In particular, an excess has been found in either the direction of local SBGs and jetted AGNs (dominated by Cen~A), over expectations for an isotropic distribution, at the $4.2\sigma$ and $3.3\sigma$ confidence level, respectively~\citep{Aab_2018, Abreu_2022}.
In \citet{TelescopeArray:2025yvu}, \auger\ and TA also did a combined analysis of the correlation between UHECR arrival directions and nearby SBGs and found a significance of $4.4\sigma$.
Even though only a small fraction (about 10\%) of the observed UHECRs correlate with the selected objects, the results motivate the study of SBGs and AGNs as UHECR sources.

Uncertainties in the UHECR mass composition, the Galactic magnetic field (GMF), and the extragalactic magnetic field (EGMF) all collectively hinder the identification of the origin of UHECRs. The role of EGMFs is that, in addition to the GMFs, they smear out the location of the UHECR source and wipe out the time correlation, which may be used to identify transients.\footnote{For example, for the AGN Cen-A, time delays of Myr are expected for EGMFs with nG strengths~\citep{2025arXiv250201022M}. The GMF adds a minimum delay of $\sim$1--50~yr for protons with $E\sim60$--100~EeV along high-latitude sightlines, and much longer ($\gtrsim10^3$~yr) for heavier nuclei. Even if extragalactic magnetic fields are negligible, this GMF “floor” limits the types of transients that can be temporally associated with UHECRs, with only the highest-energy protons along favorable directions potentially identifiable.} Understanding EGMFs is, therefore, crucial to understanding the transport of UHECRs.

Although the details about EGMFs remain unclear~\citep{Dolag_2005, Vernstrom2021}, we anticipate that measurements of the deflections of the cosmic rays from candidate sources can themselves provide new insights into the EGMF properties.  
Within local distances from their origin, before becoming isotropic, the UHECR arrival directions retain some angular correlation with their source. This residual angular structure for local sources can provide a valuable probe of the strength of the EGMF, since the extent of deflection, and hence the angular spread around the source position, is directly influenced by the EGMF strength. 
In addition to magnetic deflections, the density of UHECR sources shapes the observed angular distribution: A lower source density enhances the visibility of local sources, whose anisotropic signatures can be resolved more distinctly, whereas a higher source density leads to a more isotropic background dominated by numerous distant sources. By modeling the interplay between magnetic deflection and source distribution, one can constrain the EGMF more effectively~\citep{van_Vliet_2021}.

Combined fits to both the energy spectrum, composition, and anisotropy can provide additional insights. Such a combined analysis by \auger\ ~\citep{PierreAuger:2023htc}, using energy dependent ``blurring angles'' as a description of the effect of turbulent magnetic field deflections, found preference for a starburst origin of UHECR. Alternatively, \citet{Wykes:2017nno,Taylor:2023qdy} have found that a Cen~A origin interpretation of the energy spectrum and small-scale anisotropy is also possible within a scenario in which galaxies possess magnetized giant haloes.
A better understanding of the local EGMF is crucial here to discern between these two scenarios.

In this work, we investigate constraints on the local extragalactic magnetic field strength and coherence length on distance scales of the local sources. 
Our approach closely follows \auger 's work in~\citet{Abreu_2022} using the same methodology (see Sec.~\ref{sec:AugerAnalysis}), but explicitly incorporating the deflections caused by extragalactic magnetic fields, taking source distances into account (see Sec.~\ref{sec:ourModel}).  Consequently, in our approach, the signal fraction and magnetic field strength will be treated as key parameters. Our primary goal is to derive constraints for the EGMF strength, see Section.~\ref{sec:Results}. We will finally discuss our results and conclude in Secs.~\ref{sec:Discussion} and \ref{sec:summary}, respectively.

\section{Methods} 

This work closely follows the analysis of UHECR data by \auger\ from~\citet{Aab_2015, Aab_2018, Abreu_2022}. Throughout, we assume that the small-scale anisotropy observed by \auger\ is genuine, using the correlation with candidate sources to constrain the magnetic-field parameters required to produce these UHECR hotspots.

For the  EGMF, we adopt a model consisting of a random magnetic field characterized by a strength \( B \) and coherence length \( L_{\rm coh} \). Deflections due to the Galactic magnetic field are not included in the current analysis but will be addressed in future work. The observed anisotropy for both source catalogs is interpreted within a framework that incorporates \( B \), \( L_{\rm coh} \), and the UHECR spectrum and composition. This approach enables us to derive both lower and upper bounds on the strength of the EGMF.

\subsection{Analysis of the Pierre Auger Collaboration}
\label{sec:AugerAnalysis}

In the papers of \auger, they tested their measured arrival directions for cosmic rays with energies $E>32$~EeV against a blind search for excesses in the sky, autocorrelations, correlations with the Galactic and supergalactic planes, the Galactic center, catalogs of candidate host galaxies, and the Centaurus region. In \citep{Abreu_2022}, the entire Phase 1 data set of the Pierre Auger Observatory, i.e.~all data preceding the upgrade to AugerPrime~\citep{PierreAuger:2016qzd}, was used for this analysis. That article included a release of the data set itself, as well as the dedicated analysis software and the catalogs of galaxies that were used. The analyses in this work are based on that data set, analysis software, and those catalogs.

For the correlations with catalogs of candidate host galaxies, \auger\ considered four different catalogs. The catalog of SBGs (based on \citet{Lunardini_2019}) includes 44 sources with a radio flux at 1.4~GHz larger than 0.2~Jy from the NRAO VLA Sky Survey (NVSS)~\citep{Condon_1998A} and Parkes~\citep{Calabretta2014} surveys, and within a distance of $2{-}109$~Mpc. The catalog of jetted AGN has 26 source candidates based on the $\gamma$-ray Fermi-LAT 3FHL AGN catalog~\citep{Fermi-LAT_2017}, and with an integral flux between 10~GeV and 1~TeV larger than $3.3 \times  10^{-11} \mathrm{cm}^{-2} \mathrm{s}^{-1}$ within a distance $3{-}241$~Mpc. Besides these two catalogs, \auger\ also considered a catalog representing the large-scale distribution of matter (2MASS~\citep{2MASS:2006qir}) and a catalog of all different types of AGN (based on \citet{Oh:2018wzc}). These catalogs were used to create model maps of UHECR arrival directions.

The skymap models generated by \auger\ (see Fig.~10 of \citet{Abreu_2022}) illustrate the expected distribution of the UHECR flux across the sky, based on the cataloged source populations. These models represent the anisotropic distribution of cosmic-ray sources as inferred from the source catalogs. The models are for a cosmic-ray flux originating from the sources listed in the catalogs, with no contribution from a random, isotropic background. 

To create the model maps based on the source catalogs, the deflection of cosmic rays in Galactic and extragalactic magnetic fields needs to be taken into account. \auger\ did this by smearing out the flux distribution of the sources using Fisher distributions centered on the source positions. This is mathematically equivalent to a Gaussian distribution on the sphere~\citep{Fisher1953}. These Fisher distributions are represented by a Fisher smearing angle or search radius $\Theta$. The value of $\Theta$ is the same for all sources in a particular model map, and this value is scanned over to find the best fit to the UHECR data. Effectively, this means that \auger\ ignored the distance to the sources when estimating the deflection from the sources, implying negligible EGMF deflections and dominant GMF deflections. 
Our analysis deviates from this approach. Instead of applying Fisher distributions, we simulated the propagation of UHECRs through turbulent EGMF for different magnetic-field strengths. This allows us to obtain an upper limit on the strength of EGMFs when neglecting GMF deflections.

In the creation of the model maps, the UHECR contribution from each source is weighted according to the radio flux of the sources for the SBGs catalog and the $\gamma$-ray flux of the sources for the jetted AGN catalog. An additional distance- and energy-dependent weight is applied to each source representing the attenuation of UHECRs due to energy losses. The model maps, therefore, also depend on the threshold energy applied to the data, but only due to the energy-dependent attenuation weight. This is different in our model maps, as deflections in magnetic fields decrease with increasing energy; higher-energy events are expected to preserve more directional information about their sources. Therefore, different energy thresholds yield different model maps when magnetic-field deflections are included, with higher thresholds resulting in more anisotropic structures. 

Subsequently, the signal fraction \(\alpha\) is introduced as a key parameter in the modeling of UHECR arrival directions. It quantifies the fraction of observed events that are attributed to the astrophysical sources listed in a given catalog, with the remainder assumed to arise from an isotropic background or from sources not included in the catalog. A value of \(\alpha = 1\) corresponds to the scenario in which all events originate from the catalog sources, while \(\alpha = 0\) represents a fully isotropic distribution. Intermediate values reflect a mixed composition of catalog-associated and background events.

Next, the model flux maps are modulated by the directional exposure of the Pierre Auger Observatory, accounting for its non-uniform sky coverage. An unbinned maximum likelihood ratio is used to quantify how well the model describes the data measured by \auger . The likelihood (\(L\)) is the sum of the product of all directions of UHECR events with the model density in the same directions. The likelihood of the isotropic model (null hypothesis), denoted as \(L_0\), is the product of UHECR events with a flat probability density map, multiplied by the directional exposure of \auger . The test statistic, \(\mathrm{TS} = 2 \ln(L/L_0)\), quantifies the deviation from isotropy and serves as a likelihood ratio test between two nested hypotheses: the catalog-based model and the isotropic model. The TS is computed for each combination of the signal fraction $\alpha$, angular scale $\Theta$, and energy threshold \(E_{\mathrm{th}}\). Local p-values are then converted into global p-values.

In \auger's analysis, the catalog of SBGs gave a post-trial deviation at the $4.2\sigma$ level, while the catalog of jetted AGN gave a deviation of $3.3\sigma$. These are the two catalogs that we are considering in this work. The 2MASS catalog and the catalog of all different types of AGN gave deviations from isotropy similar to the ones from the jetted AGN catalog, while containing a significantly larger number of sources (making our analysis methods computationally unviable).

\subsection{Modeling cosmic ray deflections}
\label{sec:ourModel}

In our analysis, we construct a probability density map following an approach inspired by \auger , with one key difference: we explicitly simulate the expected deflections of UHECRs from their original source positions. We explicitly incorporate the magnetic-field strength and source distances to describe the deflections of UHECRs. In contrast, \auger\ constructs their sky model using a symmetric Fisher distribution centered on each source's position. 

\subsubsection{extragalactic magnetic field model}
\label{sec:EGMF}

We model the EGMF as a purely turbulent field, without any large-scale or structured component. 
The turbulent magnetic field in our simulations is modeled with a Kolmogorov power spectrum, with a minimum and maximum turbulence scale of $\lambda_{\min} = 120~\text{kpc}$ and $\lambda_{\max} = 1~\text{Mpc}$, respectively, resulting in a coherence length of $l_{\mathrm{coh}} = 0.256~\text{Mpc}$.
$\lambda_{\max}$ is chosen so that the resulting coherence length is significantly smaller than the typical distance to nearby sources, ensuring that UHECRs experience multiple independent deflections \citep{Batista_2019}. $\lambda_{\min}$ is chosen so that the lowest energy particles considered still have resonant modes to interact with, even for the strongest magnetic fields. In the simulations, the magnetic-field modes are assigned random directions, while the magnitude of the field strength in each mode follows a power law as described in~\citet{1999ApJ...520..204G}. We conduct our simulations for different magnetic-field strengths, sampled on a logarithmic scale, using the values listed in Table~\ref{tab:magnetic_fields}.\footnote{Table 1 gives $B_{\rm rms}$ rather than $\tilde{B}$ because $B_{\rm rms}$ is the direct input parameter used in the CRPropa simulations. The magnetic-field strengths were sampled on a logarithmic scale, with a finer step size in the middle of the parameter range to capture the transition region. While $B_{\rm rms}$ provides a clear, model-independent reference, $\tilde{B}$ is the effective quantity that directly impacts UHECR propagation and is discussed in the results.}

\begin{table}
\centering

\begin{tabular}{cccccccc}
\hline
\multicolumn{8}{c}{$B_{\mathrm{rms}}$ [nG]} \\
\hline
\hline
1.0 & 2.0 & 2.8 & 4.0 & 4.8 & 5.7 & 6.7 & 8.0 \\
8.7 & 9.5 & 11.3 & 12.3 & 13.5 & 16.0 & 32.0 & \\
\hline
\end{tabular}
\caption{Magnetic-field strengths used in the simulations, sampled on a logarithmic scale with a smaller logarithmic step size in the middle of the parameter range.}
\label{tab:magnetic_fields}
\end{table}

The characteristics of the magnetic field are parametrized by its root-mean-square strength ($B_{\mathrm{rms}}$) and maximum correlation length ($l_{\mathrm{coh}}$). These are combined into an effective parameter, $\tilde{B} = B_{\mathrm{rms}} \sqrt{l_{\mathrm{coh}}}$, consistent with theoretical expectations for cosmic ray diffusion in turbulent fields~\citep{Neronov_2010}. 

The choice of magnetic-field parameters is consistent with bounds derived from gamma-ray observations, which constrain the root-mean-square field strength to $B_{\mathrm{rms}}  \geq 10^{-17}$~G for $l_{\mathrm{coh}} \geq 0.1$~Mpc  \citep[see e.g.][]{MAGIC2023}. The selected range of $\tilde{B}$ spans from scenarios with small deflections to those with significant deflections, for a typical source distance of $\sim 4$~Mpc.


\subsubsection{UHECR propagation from the sources}
\label{sec:Propagation}

The deflection angles are computed using Monte Carlo simulations with CRPropa~3~\citep{AlvesBatista_2022}, which propagates UHECRs through the magnetic field model described in Section.~\ref{sec:EGMF}. Importantly, in our setup, the simulated deflection angle depends only on the energy of the particle and the comoving distance to the source, and not on the actual position of the source in the sky.

We simulated $5 \times 10^5$ particles per source using \texttt{CRPropa}, propagating them through our turbulent extragalactic magnetic field setup. The source sends out particles isotropically. We track all particles out to a sphere centered at the source, with a radius equal to the source distance. We evaluate the deflection angles through the comparison of the emission and observer momentum vectors. All particles arriving at the observer sphere are used to create the image for one specific source. In this way, we effectively average over the turbulent magnetic field structure orientation. Our approach ignores any coherent shifts produced by the turbulent field realization in a particular direction~\citep{Dolgikh:2025bac}. 
This “large observer” approach mitigates the extremely low detection probability of a point-like observer while preserving the directional information relevant for subsequent analysis.

To simulate the extragalactic propagation of UHECRs and their interactions en route from astrophysical sources to Earth, we account for both magnetic deflections and energy losses during transit. A comparison between the simulation results and analytical estimates for the expected UHECR deflections is presented in Appendix~\ref{App1}.

To recover the correct energy spectrum for each source, it is important to note that the threshold energy defines only the minimum energy considered. For distant sources, the observed CR spectrum becomes increasingly softer, as the highest-energy CRs undergo substantial energy losses over cosmological distances. Moreover, the magnetic field strength plays a key role by increasing the effective path length of CRs through scattering. This effect is energy-dependent: lower-energy CRs are more strongly deflected and follow longer, more tortuous trajectories, thereby incurring greater energy losses compared to their higher-energy counterparts.

We perform fully 3D simulations, conducting separate runs for each source in both catalogs. The energy distribution at the sources is given by the following:

  \be \label{eq:source_spectrum}
	\frac{{\rm d}N_{\rm i}}{{\rm d}E} \propto  
           \left\{
			\begin{array}{rcl}
          E^{\rm \gamma} \;\;\;\;\;\;\;\;\;\;\; \;\;\;\;\;\;\;\;\;\;\;\;\;\; \; \;\;\;\;\;\;\;\;\;\;\;\; & ; & E < Ze R_\mathrm{max}  \; , \\
          & & \\
            E^{\rm \gamma}  {\rm exp} \left(1- \frac{E}{ZeR_\mathrm{max}}\right)  \;\;\;\;\;\;\;\;\;\;\;& ; & E \geq Ze R_\mathrm{max} \; .
			\end{array}
			\right. 
\ee
where $E$ and $Z$ are the energy of the cosmic ray and  its charge at the source, respectively. $R_\mathrm{max}$ is the maximum rigidity at the source, $\gamma$ is the spectral index.
In this work, we use the best-fit parameters for star-formation rate source evolution obtained from \citet{Batista_2019} for $\gamma$ and $R_\mathrm{max}$ as: $\gamma =-1.3$, $R_\mathrm{max}= 10^{18.2}$~V. The source evolution itself will not have a significant effect on our simulations, as we are only simulating the arrival distribution from relatively nearby sources. For the composition, we only include nitrogen nuclei at the source as they typically provide the dominant contribution in the relevant energy range (between 32~EeV and 80~EeV), see Section.~\ref{sec:Discussion} for further discussion and justification of these assumptions.

 In our simulations, we account for all relevant propagation effects for all relevant isotopes, including: pion production, photodisintegration and pair production in ambient low-energy photon fields (the cosmic microwave background (CMB) and the extragalactic background light (EBL) from~\citet{Gilmore_2012}), as well as adiabatic energy losses and nuclear decay.

\subsection{Model map creation and data comparison}
\label{Sec:dataComparison}

In this analysis, to construct the image map for each single particle, we rotate its simulated deflection angle around its original direction vector. This generates a ring centered at the original source position with a radius equal to the simulated deflection angle. In this way, we gain in efficiency as each simulated particle generates a ring instead of a single position in the sky. A source image is created by summing the rings, each normalized to its corresponding surface, of all simulated particles. The resultant image is approximately Gaussian around the source position. 
The combination of all source images, normalized according to their fluxes in the source catalog, gives the full simulated catalogue sky map. This is created on a HEALPix map using NSIDE = 64, which corresponds to a pixel size on the order of the angular resolution of the Observatory~\citep{Górski_2005}.

In the likelihood analysis,  the signal fraction \(\alpha\) is varied along with the effective magnetic field strength \(\tilde{B}\), which governs the angular spread of events around source positions due to magnetic deflections, to identify the combination of parameters that best fits the observed UHECR arrival directions.

The test statistic is maximized as a function of \(\alpha\), in steps of \(1\%\), and the effective magnetic field strength \(\tilde{B}\), with $l_{\mathrm{coh}} = 0.256~\text{Mpc}$ and magnetic-field strengths as listed in Table~\ref{tab:magnetic_fields}. We scan the energy thresholds \(E_{\mathrm{th}}\) ranging from 32 to 80~EeV. For each energy threshold, we explore the two-dimensional parameter space defined by the magnetic field strength and the signal fraction.

For a given energy threshold, the test statistic follows a \(\chi^2\) distribution with two degrees of freedom, as described by \citet{Wilks}. Consequently, the \(1\sigma\) and \(2\sigma\) confidence levels (C.L.s) on the best-fit parameters, namely the signal fraction \(\alpha\) and the effective magnetic field strength \(\tilde{B}\), are determined from iso-TS (constant test statistic) contours in the two-dimensional parameter space. These contours correspond to regions where the TS value is reduced by fixed amounts relative to its maximum.

Specifically, the $1\sigma$ and $2\sigma$ confidence regions correspond to TS decreases of 2.3 and 6.2, respectively, reflecting joint confidence regions for two parameters rather than the one-parameter intervals (TS decreases of 1 and 4) typically used for marginalized or profiled distributions, consistent with the expectations from the $\chi^2$ distribution for two parameters~\citep{Abreu_2022}.


\section{Results of the simulations}
\label{sec:Results}

We present the results of the parameter scans over the magnetic-field strength, signal fraction, and threshold energies and for the two tested source catalogs. The results encompass various analyses, starting with the creation of probability density maps for the two catalogs. These maps serve as the basis for the next step, where we perform a maximum likelihood ratio analysis to evaluate the consistency of the simulated UHECR arrival directions with the observational data. The simulated UHECR sky maps that most closely resemble the UHECR data, and their corresponding parameter values, are shown.

\subsection{Probability density map}
\label{sec:ProbabilityMap}


\begin{figure*}
	\centering
	\includegraphics [width=\textwidth]{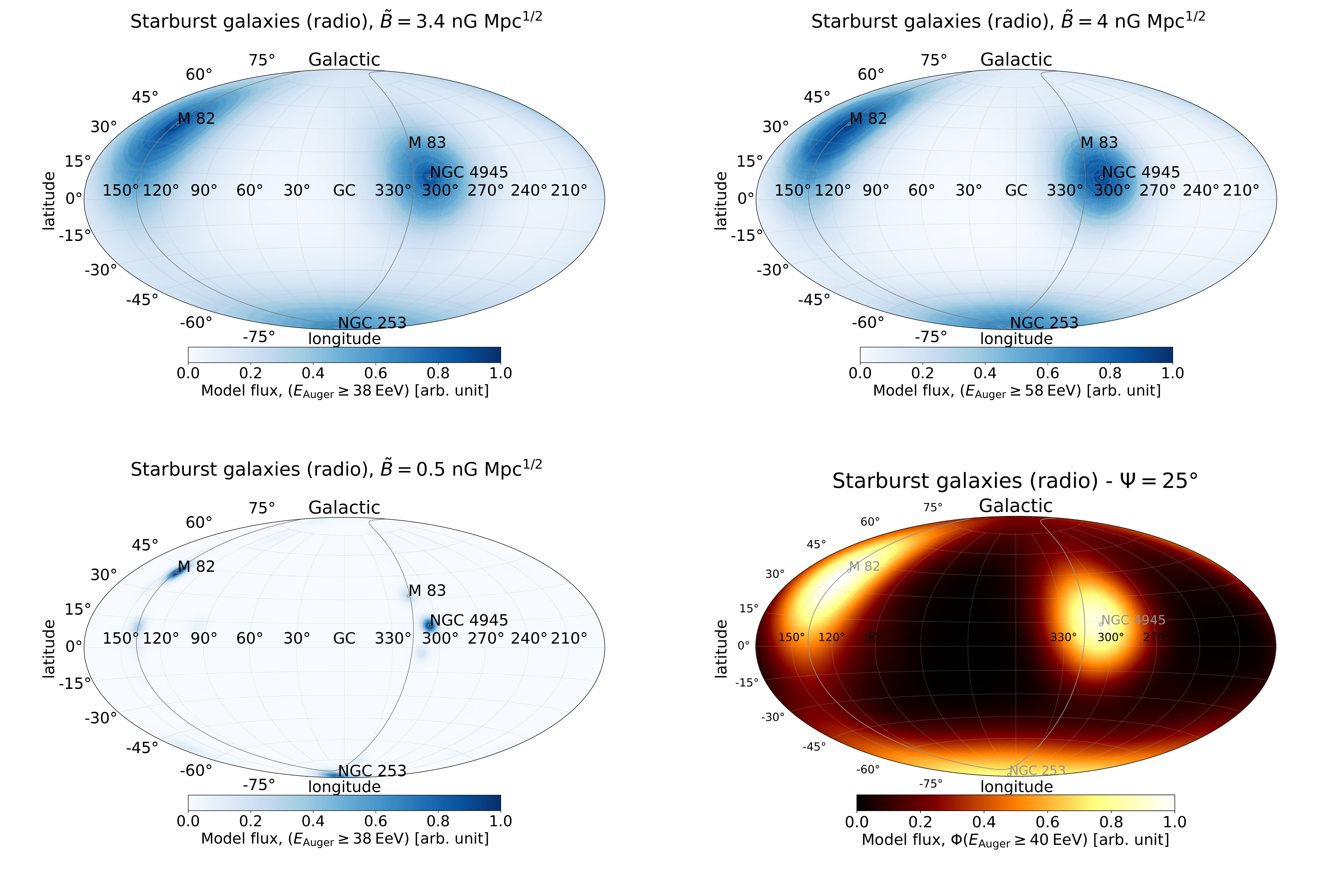}
	\caption{Probability density map for the SBG catalog in Galactic coordinates. The best fit was found at $\tilde{B}=\; 3.4$~nG~Mpc$^{1/2}$ and $E_\mathrm{th} = 38$~EeV (top left), and the secondary maximum at $\tilde{B}=\; 4$~nG~Mpc$^{1/2}$ and $E_\mathrm{th} = 58$~EeV (top right). The map (bottom left) for $\tilde{B}= 0.5$~nG~Mpc$^{1/2}$ and $E_\mathrm{th} = 38$~EeV is added as an additional example to show the effect of a reduced magnetic-field strength. The sky map in the bottom-right panel is reproduced from \citet{Abreu_2022} without modification for comparison purposes (published under the \href{https://creativecommons.org/licenses/by/4.0/}{Creative Commons Attribution 4.0 International License}). The supergalactic plane is shown as a gray line in all sky maps. Our model maps are depicted in blue to distinguish between the \auger\ map and our models. These maps do not include the contribution from the isotropic background or the exposure of \auger , which are taken into account in the TS calculations.
    } 
	\label{fig:ProbabilityMap-SBG}
\end{figure*}
\begin{figure*}
	\centering
	\includegraphics [width=\textwidth]{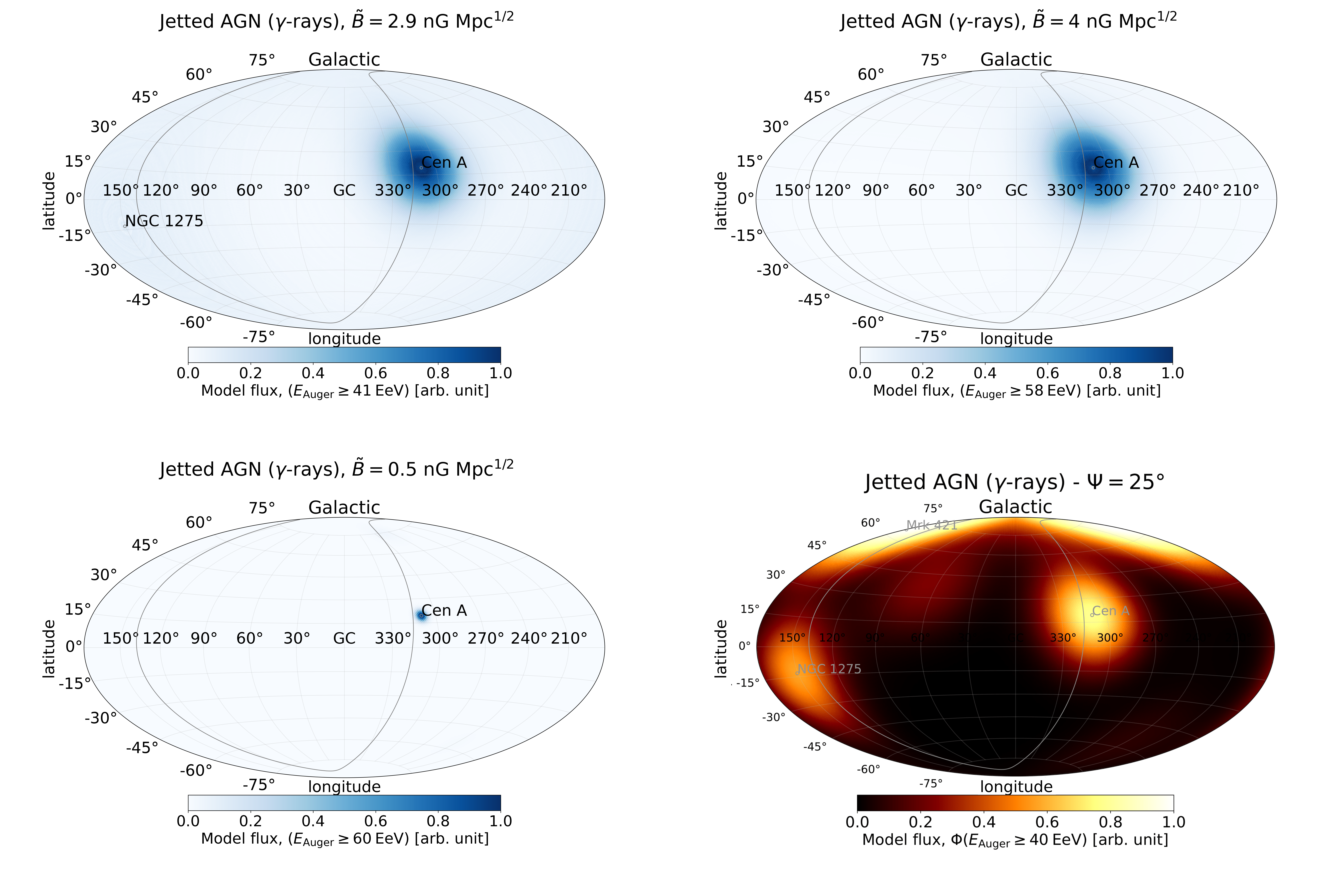}
	\caption{Probability density maps for the jetted AGN catalog in Galactic coordinates. The best fit was found at $\tilde{B} = 2.9$~nG~Mpc$^{1/2}$ and $E_\mathrm{th} = 41$~EeV (top left), and the secondary maximum at $\tilde{B} = 4$~nG~Mpc$^{1/2}$ and $E_\mathrm{th} = 58$~EeV (top right). The map (bottom left) for $\tilde{B}= 0.5$~nG~Mpc$^{1/2}$ and $E_\mathrm{th} = 60$~EeV is added as an additional example. The sky map in the bottom right panel is reproduced from \citet{Abreu_2022}  without modification for comparison purposes (published under the \href{https://creativecommons.org/licenses/by/4.0/}{Creative Commons Attribution 4.0 International License}). The supergalactic plane is shown as a gray line in all sky maps. These maps do not include the contribution from the isotropic background or the exposure of \auger , which are taken into account in the TS calculations.} 
	\label{fig:ProbabilityMap-AGN}
\end{figure*}

Probability density maps for UHECRs are essential tools in astrophysical research, as they facilitate the analysis of the spatial distribution of UHECR sources across the sky. These maps quantify the likelihood of detecting UHECR events from specific regions. Fig.~\ref{fig:ProbabilityMap-SBG} presents four probability density maps for the SBG catalog in Galactic coordinates. These maps illustrate the expected spatial distribution of UHECR arrival directions, as determined by the contribution of the sources in the SBG catalog.

To observe the effect of deflection in the EGMF, we plot the probability density map at different magnetic-field strengths.
In Fig.~\ref{fig:ProbabilityMap-SBG}, we show the distribution of the best-fit sky map in Galactic coordinates for the SBG sources with $\tilde{B}= 3.4$~nG~Mpc$^{1/2}$ and $E_\mathrm{th} = 38$~EeV (top left). It demonstrates a localized clustering of the cosmic-ray flux, with visible deflection effects caused by the EGMF. This map can be compared with the best-fit model map of \auger\ (bottom right) above $40$~EeV, without modification~\citep{Abreu_2022}. This map of \auger\ does not account for distance-dependent deflection effects, but its distribution is similar to our maps as the dominant sources, M~82, NGC~4945, and NGC~253, are all at around 3.5 to 3.7~Mpc distance.

The top-right panel corresponds to a magnetic-field strength of $\tilde{B}= 4$~nG~Mpc$^{1/2}$ at $E_\mathrm{th} = 58$~EeV, which is a secondary local maximum in the parameter space. The clustering in this case is similar due to a stronger magnetic-field strength (leading to stronger deflections) and an increase in the energy threshold (decreasing the deflections), but is compared with a different data set, as the threshold energy is applied to the observed data as well. The bottom-left panel shows the reduction of deflections for a smaller magnetic-field strength, at $\tilde{B}= 0.5$ nG Mpc$^{1/2}$ and $E_\mathrm{th} = 38$~EeV.

For all panels shown in Fig.~\ref{fig:ProbabilityMap-SBG}, the supergalactic plane is marked as a gray line in all panels, serving as a reference for identifying large-scale structures. As shown, the expected anisotropies in the arrival directions of SBGs are dominated by the same sources that contribute most significantly in the \auger\ analysis. However, notable differences exist between our model and the one used by \auger . In our approach, we explicitly account for deflections in the magnetic field as well as the distances to the sources. For nearby sources, magnetic deflections are minimal, resulting in a higher concentration of arrival directions near the actual source position.

In contrast to our approach, the \auger\ methodology assigned the same angular width of the Fisher distribution to all sources in the catalog, irrespective of the distance to the source. In particular, in our approach, the closest sources are also the brightest, forming a hot spot expected in the \auger\ field of view, particularly in the direction of the group of galaxies composed of NGC~4945, NGC~253, M~82, and Centaurus~A (see Fig.~\ref{fig:ProbabilityMap-AGN}). All sources depicted in the map are very luminous and are located at distances of $< 4$~Mpc from Earth. The M~82 hotspot lies mostly outside the exposure region of the \auger , which makes it clearly observable by TA. The deflection caused by the magnetic field is minimal for small magnetic fields but increases as we raise the magnetic-field strength, as indicated in the panels.

Fig.~\ref{fig:ProbabilityMap-AGN} focuses on the jetted AGN catalog and similarly shows the distribution of the best-fit sky map, the secondary maximum, the \auger\ best-fit, and an additional example map, all in Galactic coordinates. The best-fit map is shown in the top-left panel, with $\tilde{B} = 2.9$~nG~Mpc$^{1/2}$ and $E_\mathrm{th} = 41$~EeV. This map highlights the concentration of flux near Centaurus~A, the dominant source in this catalog, reflecting its strong contribution to the observed anisotropy. The top-right panel of Fig.~\ref{fig:ProbabilityMap-AGN} shows the secondary local maximum, with $\tilde{B} = 4$~nG~Mpc$^{1/2}$ and $E_\mathrm{th} = 58$~EeV. 
The clustering in these two cases is again similar due to a stronger magnetic field strength and an increase in the energy threshold. The bottom-left panel shows the probability density map for $\tilde{B}= 0.5$ nG Mpc$^{1/2}$ and $E_\mathrm{th} = 60$~EeV  to show the effect of the EGMF strength on the UHECR distribution.

The sky map on the bottom right panel of Fig.~\ref{fig:ProbabilityMap-AGN} is taken from \citet{Abreu_2022} for comparison with our model maps. \auger 's map is significantly different from our model maps, as, in our case, the expected anisotropies in the arrival directions of jetted AGNs are dominated by Centaurus~A alone, in contrast to the \auger\ analysis, which identifies significant contributions from Mrk~421 and NGC~1275 as well. The difference is because, in our maps, the UHECRs from distant sources are strongly deflected, while in \auger 's map, the spread around the source position is the same no matter what the distance to the source is. With distances for Cen~A of about 3.7~Mpc, NGC~1275 of about 78~Mpc, and Mrk~421 of about 134~Mpc, this results in large spreads (increased by significant energy losses) in our models for NGC~1275 and Mrk~421. 

In both the SBG and AGN maps, incorporating source distance and magnetic deflections provides a more nuanced understanding of UHECR propagation and clustering effects. The proximity and luminosity of sources such as Centaurus~A dominate the anisotropic signal, consistent with observational data. In our model, the distance to the source is explicitly accounted for in the calculations. This ensures that the observed flux accurately reflects the true luminosity of the source, taking into consideration its spatial position.

However, in \auger 's maps, the source distance is not directly observable. This discrepancy arises because, in \auger 's maps, the source-flux weight effectively compensates for the change in intensity due to the source's distance, which results in the distance not being explicitly visible in the final map. In this way, the map may present integrated or averaged flux values that obscure the individual contributions from sources at different distances. This approach ensures that the map remains focused on the intensity or brightness distribution without highlighting spatial variations attributable to varying distances.


\subsection{Maximum likelihood ratio analysis}
\label{sec:Maximumlikelihood}

Using the unbinned maximum likelihood method described in Section~\ref{Sec:dataComparison}, we estimate the best-fit signal fraction $\alpha$ and effective magnetic field strength $\tilde{B}$ for each energy threshold.
This allows us to evaluate how the correlation between UHECR arrival directions and catalog sources evolves with energy.

\begin{figure*}
	\centering
	\includegraphics [width=\textwidth]{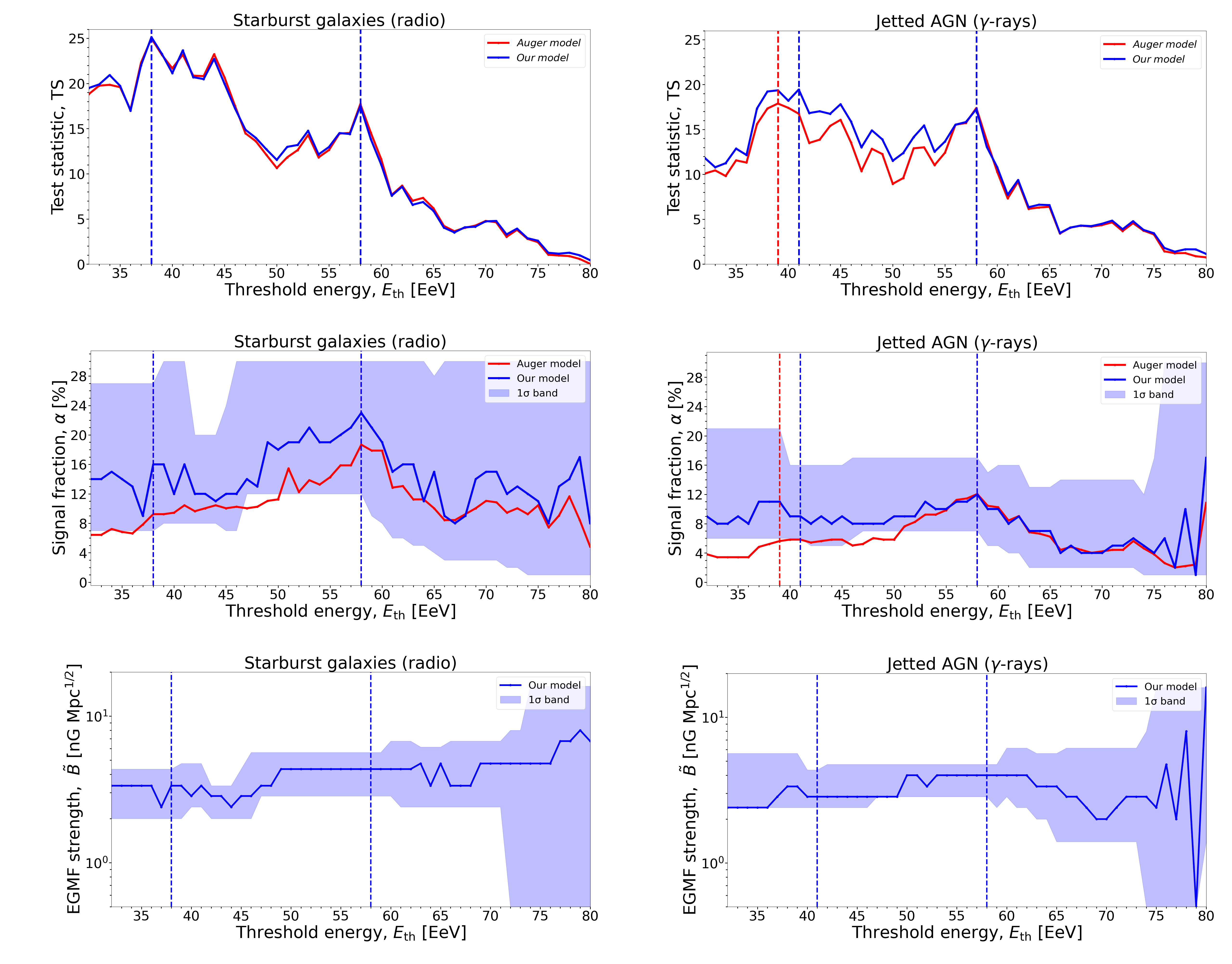}
	\caption{The test statistic (top), signal fraction (middle), and magnetic-field strength (bottom) when maximizing the deviation from isotropy for the covered range of threshold energies, for SBGs (left), and jetted AGNs (right). The results of our model (blue) are compared with the results from~\citet{Abreu_2022} (red) for the test statistic and corresponding value of the signal fraction. The vertical lines indicate the positions of the best fit and the secondary maximum.   The 1$\sigma$ confidence intervals for $\alpha$ and $\tilde{B}$ shown with blue shading, are obtained using all TS maxima with energies above the threshold. The TS curve exhibits a double-peak structure: the first peak corresponds to the global TS maximum, while the second peak contains fewer contributing events resulting in lower statistical significance.}
    \label{fig:Maximum-likelihood}
\end{figure*}

Fig.~\ref{fig:Maximum-likelihood} shows the values of $\alpha$ and $\tilde{B}$ that maximize the TS for each energy threshold. Results are presented separately for the SBG and jetted AGN catalogs. The results for SBGs and AGNs from our model are depicted in blue, while the corresponding results from the \auger\ analysis are shown in red. This figure provides insight into how the signal fraction and inferred magnetic deflection scale with energy, and which catalog yields a better fit to the observed UHECR distribution. 

As illustrated in Fig.~\ref{fig:Maximum-likelihood}, the TS curve reveals a two-peak structure as a function of the energy threshold (top panel). For SBGs (left), the best-fit peak is at a threshold energy of 39~EeV, while for AGNs (right) it is at 41~EeV. The second peak, at energies of 58~EeV for both catalogs, is consistent with the findings reported by \auger.

The first peak (39~EeV) corresponds to the global maximum of the TS. The second peak (58~EeV) coincides with the maximum inferred signal fraction for the catalogs but involves fewer events, resulting in lower statistical significance. Given that the 1$\sigma$ uncertainty on the TS is approximately 2.3, the secondary peak may result from statistical fluctuations rather than genuine signals. The observed double-peak structure therefore reflects the interplay between the number of events and the inferred signal fraction, with only the global maxima considered robust.

In the middle panel of Fig.~\ref{fig:Maximum-likelihood}, we compare the optimal values of the signal fraction $\alpha$ across the energy thresholds for our model (blue) against those of the \auger\ results (red).
The final panel illustrates the magnetic field strength $\tilde{B}$ values that correspond to the maximized TS. The variation illustrates how the best-fit magnetic-field strength evolves across the scanned energy thresholds.

In table~\ref{table:1}, we show the values of $E_\mathrm{th}$, $\alpha$, and $\tilde{B}$ and the TS for the global maximum and the secondary local maximum, for the SBG and jetted AGN catalogs.

\begin{figure*}
	\centering
	\includegraphics [width=\textwidth]{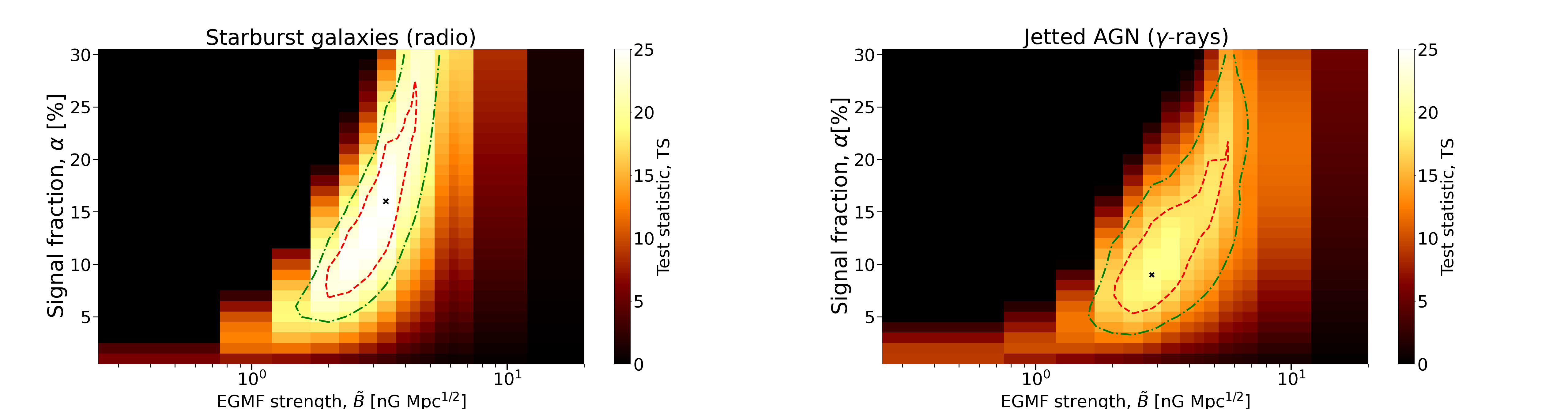}
	  \caption{The test statistic values for SBGs (left) and jetted AGNs (right) as a function of the signal fraction and magnetic-field strength, evaluated at a threshold energy of 38~EeV for SBGs and 41~EeV for jetted AGNs. The dashed green and red lines represent $68.3\%$ and $95.4\%$ C.L.~contours (1~d.o.f.), respectively. The crosses indicate the best-fit positions.}

	\label{fig:TS-SignalFraction}
\end{figure*}

To further investigate the behavior of the likelihood landscape and assess the uncertainties on the model parameters, we perform a comprehensive 2D scan over the signal fraction $\alpha$ and magnetic field strength $\tilde{B}$ at the energy threshold where the test statistic reaches its maximum value. This scan allows us to visualize how the test statistic varies across the parameter space and to identify regions that provide the best agreement between the observed UHECR distribution and the model predictions based on each source catalog. By examining the shape and extent of the confidence level contours, we gain insight into the stability and degeneracy of the fit parameters, as well as the statistical preference for specific combinations of $\alpha$ and $\tilde{B}$. The results for both SBG and AGN catalogs are shown in Fig.~\ref{fig:TS-SignalFraction}.

In the left panel of Fig.~\ref{fig:TS-SignalFraction}, the results for the SBG catalog are shown. This plot highlights the region in the $(\alpha, \tilde{B})$ plane where the TS reaches its maximum. The dashed red and green contours correspond to the $68.3\%$ and $95.4\%$~C.L., respectively, derived from the TS distribution under the Wilks’ theorem approximation. The right panel presents the corresponding results for the jetted AGN catalog. As in the left panel, the best-fit parameters are marked with a cross, and the C.L. contours illustrate the uncertainties on the fit parameters. The best-fit parameters obtained above the energy threshold that maximizes the departure from isotropy are marked with a cross.

\begin{table*}
\centering
\begin{tabular}{|p{3.2cm}|c|c|c|c|}
\hline
\textbf{Catalog / Maximum} & \textbf{$E_\mathrm{th}$ [EeV]} & \textbf{Signal fraction ($\alpha$) [$\%$]} & \textbf{$\tilde{B}$ [nG Mpc$^{1/2}$]} & \textbf{TS} \\
\hline
\hline
\multicolumn{5}{|c|}{\textbf{SBGs}} \\
\hline
Global maximum    & 38 & $16^{+11}_{-9}$ & $4^{+1}_{-2}$ & 25.13 \\
Secondary maximum & 58 & $19^{+10}_{-11}$ & $4^{+3}_{-2}$ & 17.50 \\
\hline
\multicolumn{5}{|c|}{\textbf{Jetted AGNs}} \\
\hline
Global maximum    & 41 & $9^{+12}_{-3}$  & $3^{+3}_{-1}$ & 19.45 \\
Secondary maximum & 58 & $11^{+11}_{-5}$ & $4^{+2}_{-2}$ & 17.40 \\
\hline
\end{tabular}
\caption{The values of $E_\mathrm{th}$, $\alpha$, and $\tilde{B}$ obtained for the SBG and jetted AGN catalogs at the global and secondary (lower) maxima. The $1\sigma$ error ranges, following from the TS calculations in the parameter scan, are indicated for $\alpha$ and $\tilde{B}$.}
\label{table:1}
\end{table*}
From the two-dimensional likelihood scans presented in Fig~\ref{fig:TS-SignalFraction}, we also derive upper limits on the effective magnetic field strength $\tilde{B}$. By profiling the likelihood over the signal fraction $\alpha$ and identifying the boundary at which the test statistic decreases by 2.71 units from its maximum (corresponding to a 90\% confidence level for one degree of freedom), we place constraints on the maximum allowable deflections consistent with the observed data. The resulting upper limits on the magnetic field strength are $\tilde{B} < 4.4$~nG~Mpc$^{1/2}$ for the SBG catalog and $\tilde{B} < 6.7$~nG~Mpc$^{1/2}$ for the jetted AGN catalog, both at the 90\% confidence level.
The resulting lower limits on the magnetic field strength are $\tilde{B} > 2$~nG~Mpc$^{1/2}$ for both the SBG and jetted AGN catalogs, at the 90\% confidence level.
The EGMF strength limits obtained provide valuable insight into the extragalactic magnetic environment and the degree of deflection experienced by UHECRs en route to Earth.

\section{Discussion}
\label{sec:Discussion} 

In this work, we have focused on the deflections that UHECRs incur during their propagation in the EGMF. Additional deflections of UHECRs, however, are expected in the GMF. The GMF consists of coherent and turbulent magnetic field components. Including the turbulent GMF would only have further increased the rms average deflection angle. Our results, therefore, provide robust upper limits on the EGMF strength \citep[see e.g.][]{Bray:2018}. However, the lower limits on the EGMF are only applicable if the deflections in the GMF are negligible. The expected deflections in the GMF differ significantly between different GMF models. This can, for example, be seen from~\citet{Unger:2025xkk} where expected deflections in the GMF are shown for eight different variations of the UF23 models~\citep{unger2024}. Furthermore, should a magnetized Galactic halo is present around our Galaxy at larger scales than in the UF23 models, the expected deflections in the GMF increase ~\citep [see e.g.][]{Shaw:2022lqd}.   

Although we do not explicitly consider the deflection introduced by the GMF, our EGMF upper limit can be re-normalized to give correspondingly larger upper limits on the turbulent GMF. The level of this re-normalization is dependent on the propagation distance through the turbulent GMF (see Eqn.~\ref{eq:theta_rms_approx}). As an example, for a Galactic halo of size $D=30$~kpc, and a coherence length of $l_{\rm coh}=1$~kpc, an EGMF upper limit of 4~nG obtained for sources with $D\approx 3.6$~Mpc would convert to an upper limit constraint on the turbulent GMF of $1.4~\mu$G~\footnote{Interestingly, for the case in which the UHECRs arriving to Earth are dominated by deflections in the Galactic halo, the deflection angles for particles with energies below 10~EeV would expect $\theta_{\rm rms}> 100^{\circ}$, resulting in the observed skymap becoming dipolar-like, see~\citet{Shaw:2025ykm}.}. This limit sits close to, and is compatible with, the present detection hints of the halo GMF~\citep{Kronberg:2007dy,Bernet:2008qp,Heesen_2023}.

Our EGMF setup only includes turbulent EGMFs; no structured EGMFs were considered. Turbulent magnetic fields typically cause a spreading out of arrival directions around the original source position, 
provided the distance traveled is significantly larger than the coherence length  \citep{Dolgikh:2025bac}.

 Structured magnetic fields typically do the opposite; they change the apparent source position but do not cause a spread of arrival directions around the apparent source position (see e.g.~\citet{Shaw:2022lqd} for a discussion on these two effects). Our results, therefore, hold specifically for turbulent EGMFs with a maximum correlation length less than the typical source distances (3 to 4~Mpc for the nearest sources we consider). When comparing the overdensities in the UHECR arrival directions of the \auger\ data~\citep{Abreu_2022} with the source positions by eye, a shift of the source positions does not seem to be necessary if the overdensities are dominated by the nearest sources in the catalogs within the field of view of the \auger\ (M~83, NGC~4945, Circinus, and NGC~253 for starburst galaxies, and Cen~A for jetted AGN). However, for the jetted AGN case, the overdensity in the Galactic South is hard to explain with turbulent magnetic fields alone, which drives the higher significance for a deviation from isotropy for the starburst galaxies case over the jetted AGN case (although the preference for starburst galaxies over jetted AGN should not be considered statistically significant). Including the source distances and the related deflections in structured EGMFs carries further information which could be used for ``tomographic'' analyses if better statistics and UHECR composition measures are available in the future.

For the deflection angle simulations, we fixed the source UHECR composition to nitrogen. Additionally, the source spectral index and maximum rigidity were set to values obtained from \citet{Batista_2019}. These choices have been found to provide a good fit to the spectrum and composition measured by \auger\ in the relevant energy range. \citet{Ehlert:2022jmy} have shown that the maximum rigidity cannot vary much from source to source to maintain a good fit to the \auger\ data. The UHECR composition, however, can be heavier than pure nitrogen, especially at the high energy end of the considered energy range (depending on the choice of hadronic interaction model). For the same EGMF, a heavier composition would increase the angular deflection, resulting in stronger upper limits on the EGMF being obtained. The choice for a nitrogen composition, therefore, is conservative for obtaining upper limits on the magnetic-field strength. A predominantly lighter composition than nitrogen for $E > 32$~EeV is hard to reconcile with the \auger\ composition data. A subdominant light component might still exist, but would have to be produced by sources with a higher $R_\mathrm{max}$ than the dominant sources (assuming rigidity-dependent cutoffs). It is, therefore, not likely that this is produced by relatively nearby sources from one of the catalogs.

The interactions of UHECRs with background EBL and CMB photon fields can play a significant role for the catalog sources. For nitrogen with an energy of $E \sim 30$~EeV, the typical energy-loss-length is $\gtrsim 200$~Mpc (see e.g.~\citet{Hooper:2006tn,AlvesBatista:2015jem}). At higher energies, this loss length exponentially decreases~\citep{Taylor:2009iw}, with the loss length for $E\sim 50 ~{\rm EeV}$  becoming $\sim  70$~Mpc. These energy losses significantly reduce the potential contribution from sources like Mrk~421 (at 133.7~Mpc distance in the catalog) and NGC~1275 (at 78.0~Mpc). In \auger's analysis, these interactions are taken into account by applying energy-dependent and distance-dependent weights to the sources based on the expected attenuation. In our analysis, we use the same weights as computed by \auger\ for the normalization of each source to follow their analysis as closely as possible. However, in our simulations, the effect of energy losses is stronger for scenarios with stronger magnetic fields, as the effective propagation distance is increased when the deflections in magnetic fields become stronger, an effect of particular importance given the exponential decrease of the loss length noted above. In addition, cosmic rays with lower rigidities are more strongly deflected by magnetic fields, also effectively increasing their propagation distance. These effects introduce additional dependencies of the expected arrival directions on the magnetic-field strength and the energy threshold, which are not present in scenarios where a blurring based on the source distance is taken into account \citep[e.g.][]{PierreAuger:2023htc}.

In an earlier study, \citet{van_Vliet_2021} derived magnetic field constraints for SBGs using an independently developed foreground–background model that interpolates between cataloged sources (foreground) and an isotropic background beyond the listed sources. The local source density, $\rho_0$, was introduced—alongside the magnetic field strength, $\tilde{B}$—to parametrize the relative contribution of the isotropic background. Since $\rho_0$ can be directly compared to expectations from star-forming and spiral galaxies, robust lower limits on the magnetic field strength, $\tilde{B} \gtrsim 0.2,\mathrm{nG}$~Mpc$^{1/2}$, could be derived for SBGs ~\citep[see also][]{Durrer:2013pga} for an estimated lower limit on the EGMF strength from UHECRs). These limits are consistent with the constraints obtained in the present study. The origin of this lower limit lies in the requirement that either sufficiently strong magnetic fields must be present to produce the isotropic component of the map, or the source distribution itself must be intrinsically isotropic (i.e., isotropic background dominated).

In the current study, the reasoning behind the lower limit on $\tilde{B}$ is somewhat different. While the isotropic background still needs to be accounted for, magnetic deflections must not be too small, as this would result in correlations that are too strong compared to observations. For a larger signal fraction, larger values of $\tilde{B}$ are also permitted to prevent excessive correlations (see, e.g., Fig.~\ref{fig:TS-SignalFraction}).

Other works stating upper limits on the strength of EGMFs based on UHECR data are, e.g., \citet{Bray_2018, Bister:2023icg}; and~\citet{Neronov:2021xua}. These papers state lower magnetic field strengths than obtained in our work, but are generally based on stronger assumptions (besides using different methods). \citet{Bray_2018}, for example, based their limits on a typical source distance of around 70~Mpc, while their results become comparable to our upper limit for a scenario where they assumed a typical source distance of 10~Mpc. 
Rather than using the \auger\ catalog, which motivated the original correlation, \citet{Bister:2023icg} assumed UHECR sources following the local matter density. This assumption leads to an upper limit on the EGMF for specific source densities. In addition, they included GMF deflections in a specific GMF model, instead of neglecting the GMF deflections, to obtain their upper limits on the EGMF. 

\citet{Neronov:2021xua} obtained their upper limits on the EGMF assuming that the TA excess~\citep{TelescopeArray:2021dfb} is produced by the Perseus-Pisces supercluster at a distance of about 70~Mpc. Unfortunately, \auger\ has not been able to confirm the existence of an excess in the same direction, despite having a comparable integrated exposure in that region~\citep{PierreAuger:2024hrj}.  
We note that, even if the TA excess is real, there may exist sources much closer than 70~Mpc, such as NGC~891~\citep{Anchordoqui:2022pzo}, or even M31~\citep{Plotko:2022urd}, which could contribute significantly to the observed flux.

Our derived EGMF constraints relate to the EGMF between the (studied) local galaxies and the Milky Way. These constraints are not applicable to EGMF voids in general. The distances to the dominant local sources (M~82, NGC~4945, NGC~253, and Cen~A) are typically $< 4$~Mpc. The larger typical source distances assumed in \citet{Bray_2018, Bister:2023icg}; and~\citet{Neronov:2021xua} correspond more closely to limits on EGMFs in voids.

While our test statistics yield similar results to that of the original \auger\ analysis, we find a slight improvement of the fit for lower threshold energies for the jetted AGN catalog. This best fit is now more pronounced with respect to the secondary maximum. 
The inferred signal fractions range between about 10 and 20\% (depending on energy threshold and source class), which means that a substantial fraction of the observed UHECRs, beyond the chosen threshold energy, could be correlated with the sources. In fact, we do not obtain very strong upper limits for the signal fraction, especially for SBGs, which means that we cannot exclude that most of the UHECRs come from SBGs.

We have also demonstrated that while for SBG the same sources as for the  \auger\ analysis dominate the correlations, observable correlations with jetted AGNs other than Cen~A in the corresponding source catalog are eliminated by the EGMF once the source distances and corresponding deflections are included (similar to what was found in~\citet{PierreAuger:2023htc}). These particular sources could help to discriminate between potential Galactic and extragalactic magnetic fields or isolate the regions with the strongest deflections in the future.

The future of UHECR research looks promising with the advancements brought by AugerPrime~\citep{PierreAuger:2016qzd} and the Global Cosmic Ray Observatory (GCOS) initiative~\citep{GCOS:2021exh}. AugerPrime, an ongoing upgrade to the Pierre Auger Observatory, will significantly enhance the precision of UHECR energy and composition measurements, potentially allowing the composition to be measured on a shower-by-shower basis. Meanwhile, the GCOS project aims to establish a global network of observatories, including AugerPrime. By combining data from multiple observatories, GCOS will help identify correlations between UHECRs and astrophysical sources, such as active galactic nuclei and gamma-ray bursts. Together, AugerPrime and GCOS will push the boundaries of cosmic ray research, enabling scientists to investigate the origins, propagation, and anisotropies of UHECRs, and make strides in answering some of the most fundamental questions in astrophysics.

\section{Summary and conclusions}
\label{sec:summary}

We have here revisited a recent analysis by \auger\ which found a correlation between the UHECR arrival directions and the distribution of local SBGs or jetted AGNs~\citep{Abreu_2022}. Their analysis parameterized the deflections from EGMFs with a generic, source-distance independent ``Fisher search radius''. The arrival directions of a fraction (``signal fraction'') of the full UHECR dataset, for particles with energies above an ``energy threshold'', were found to correlate with the source catalog. The three parameters of their model (Fisher search radius, signal fraction, energy threshold) were then varied to maximize the test statistics. 

We adopted the original \auger\ model as close as possible, but have replaced the parameter ``Fisher search radius'' with a more realistic EGMF model, taking into account the individual (distance-dependent) EGMF deflections of the individual sources. Subsequently, we calculated these deflections (and the UHECR energy losses) from detailed UHECR transport simulations, considering the expected UHECR spectrum and composition in the relevant energy range. Consequently, the corresponding new parameter we constrained is the turbulent EGMF strength $\tilde B$ in nG~Mpc$^{1/2}$, where the degeneracy with the coherence length is taken into account to first order. The other two parameters (signal fraction and energy threshold) were adopted with the same meaning as in \citet{Abreu_2022}.

As our main result, we have derived a 90\% C.L.~upper limit value on the EGMF. A constraint of 4.4~nG~Mpc$^{1/2}$ is obtained for the SBGs scenario, and 6.7~nG~Mpc$^{1/2}$ for the jetted AGN scenario. The corresponding upper limit on the turbulent GMF in the Galactic halo is dependent on the assumed halo size. Adopting a halo of size 30~kpc, an upper limit of $1.4~\mu$G~kpc$^{1/2}$ is obtained.

\section*{Data availability}
The data sets generated and/or analyzed during this study are available from the corresponding author upon reasonable request.
The UHECR propagation simulations were performed using the publicly available software package CRPropa~3~\citep{AlvesBatista_2022}.
The availability of \auger 's software and data is described in~\citet{Abreu_2022} and is publicly accessible at \url{https://doi.org/10.5281/zenodo.6504276}.

\section*{Acknowledgements}
AAZ and AvV acknowledge support from Khalifa University’s internal FSU-2022-025 and RIG-2024-047 grants. AT and WW  acknowledge support from DESY (Zeuthen, Germany), a member of the Helmholtz Association HGF. We wish to acknowledge the contribution of Khalifa University's high-performance computing and research computing facilities to the results of this research.
We would like to thank the Pierre Auger Collaboration for making their data and software used for \citet{Abreu_2022} publicly available.
We would like to thank the participants of the ``Generation, evolution, and observations of cosmological magnetic fields'' scientific program at the Bernoulli Center in April 2024, and the program ``Towards a Comprehensive Model of the Galactic Magnetic Field'' at Nordita in April 2023, which is partly supported by NordForsk the Royal Astronomical Society, for useful discussions.

\bibliographystyle{mnras}
\bibliography{Al_Zetoun_references}

\appendix

\section{Comparison with analytic estimates}
\label{App1}

A comparison between the simulation results and an analytical estimate of the angular deflection of high-energy particles due to turbulent extragalactic magnetic fields provides valuable insight into the accuracy and limitations of the analytical models used for predicting UHECR deflections. A simple analytical expression for the root-mean-square (rms) deflection angle \( \theta_{\text{rms}} \) can be obtained assuming a set of consecutive small-angle deflections over its propagation distance \(D\) through a turbulent field, the rms deflection angle \(\theta_{\mathrm{rms}}\) can be approximated as \citep{Harari:2002dy,Bray_2018}:

\begin{eqnarray}
\theta_{\mathrm{rms}} = \frac{1}{2} \left(\frac{D}{l_{\mathrm{coh}}}\right)^{1/2} \frac{l_{\mathrm{coh}}}{r_L},
\label{eq:theta_rms}
\end{eqnarray}

\noindent where \(l_{\mathrm{coh}}\) is the magnetic field coherence length and \(r_L\) is the Larmor radius of the cosmic ray. Expressing the Larmor radius in terms of particle charge \(Z\), energy \(E\), and magnetic field strength perpendicular to the line of sight \(B_{\perp}\), this can be rewritten as:
\begin{figure*}
    \centering
    \includegraphics[width=0.80\textwidth]{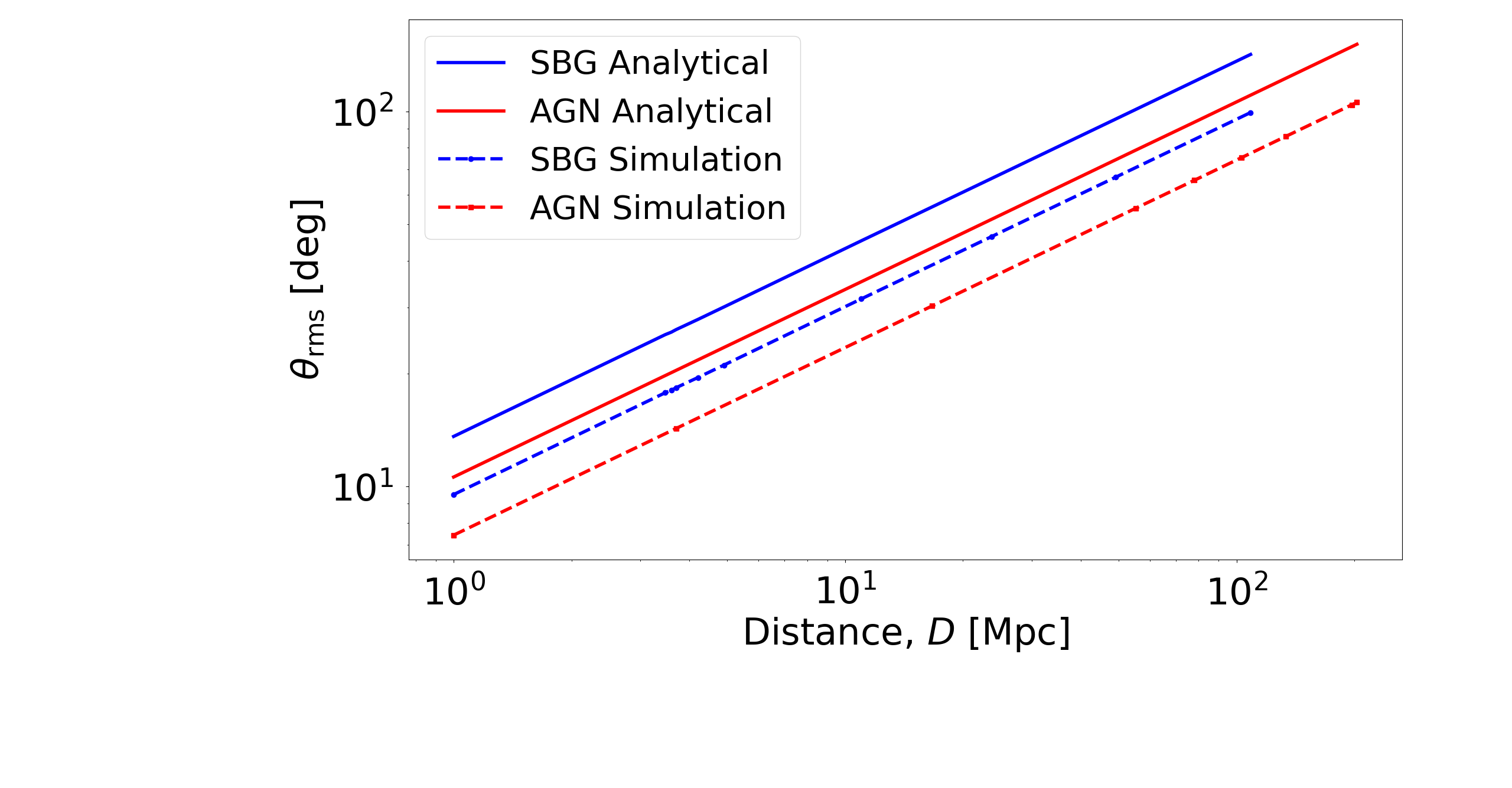} 
    \caption{Comparison of deflection angles from analytical formulae and simulation results for SBGs (blue) and AGNs (red), for an EGMF strength of 6.7\,nG and 5.7\,nG, respectively. Solid lines represent the analytical predictions, while dashed lines indicate the simulation results.}
    \label{fig:deflection_angle_comparsion}
\end{figure*}


\begin{equation}
\begin{split}
\theta_{\mathrm{rms}} \approx 8.4^\circ \times  
& \left(\frac{D}{10\,\mathrm{Mpc}}\right)^{1/2} 
\left(\frac{l_{\mathrm{coh}}}{1\,\mathrm{Mpc}}\right)^{1/2} \\
& \times \left(\frac{B_{\perp}}{10^{-9}\,\mathrm{G}}\right) 
\left(\frac{E}{eZ ~10\,\mathrm{EV}}\right)^{-1}.
\end{split}
\label{eq:theta_rms_approx}
\end{equation}

The root-mean-square magnetic field strength relates to the perpendicular component via:
\[
B_\mathrm{rms} = B_{\perp} \sqrt{\frac{3}{2}}.
\]

The result of the calculation for SBGs, assuming  

$D = 3.6~{\rm Mpc}$, $\quad B_{\mathrm{rms}} = 6.7~{\rm nG}$, $\quad l_{\mathrm{coh}} = 0.256~{\rm Mpc}$, and $E/eZ = 5.4~{\rm EV}$,
\[
\theta_{\mathrm{rms}} \approx 25.8^\circ.
\]
while the value obtained from our simulation is
\[
\theta_{\mathrm{rms}}^{\mathrm{sim}} = 19.1^\circ.
\]

For the jetted AGN catalog, assuming  
$D = 3.7~{\rm Mpc}$, $\quad B_{\mathrm{rms}} = 5.7~{\rm nG}$, $\quad l_{\mathrm{coh}} = 0.256~{\rm Mpc}$, and $E/eZ = 5.9~ {\rm EV}$, the result is:
\[
\theta_{\mathrm{rms}} \approx 20.4^\circ.
\]

while the simulation yields:
\[
\theta_{\mathrm{rms}}^{\mathrm{sim}} = 14.3^\circ.
\]

The comparison between the analytical estimates and simulation results is shown in Fig~\ref{fig:deflection_angle_comparsion}. The Figure displays the root-mean-square deflection angle \(\theta_{\mathrm{rms}}\) as a function of source distance \(D\) for SBGs (blue) and AGNs (red), assuming an extragalactic magnetic field (EGMF) strength of 6.7\,nG and 5.7\,nG, respectively.

An exact agreement between the analytical estimate and the simulation results is not expected for several reasons. For example, the analytical estimate assumes a single energy, while the simulated results cover a wide range of energies starting from a corresponding threshold energy. Furthermore, the analytical estimate assumes one particle charge, while in the simulations, the particles start with that corresponding charge, but can produce secondaries and lose charge due to interactions during the propagation.

\end{document}